\documentclass[lettersize,journal]{IEEEtran}
\usepackage{amsmath,amsfonts}
\usepackage{algorithmic}
\usepackage{algorithm}
\usepackage{array}
\usepackage[caption=false,font=normalsize,labelfont=sf,textfont=sf]{subfig}
\usepackage{textcomp}
\usepackage{stfloats}
\usepackage{url}
\usepackage{verbatim}
\usepackage{graphicx}
\usepackage{amsmath}
\usepackage{siunitx}
\usepackage{stmaryrd}
\usepackage{tikz}
\usepackage{hyperref}
\usepackage{pgfplots}
\pgfplotsset{compat=1.18}
\usepackage{booktabs}
\usepackage{upgreek}
\usetikzlibrary{patterns}

\usetikzlibrary{shapes.misc}

\newcommand{\gradient}{\nabla}
\newcommand{\Hgrad}{H^1 \! \left(\Omega_\text{c}\right)}

\newcommand{\curl}{\nabla \times}
\newcommand{\Hcurl}{H_{\phi,I} \! \left(\text{curl}, \Omega\right)}
\newcommand{\Hcurlzero}{H_{\phi,0} \! \left(\text{curl}, \Omega\right)}

\newcommand{\Hgradzero}{H^1 \! \left(\Omega_\text{c}\right)}

\newcommand{\Hcurlnophi}{H \! \left(\text{curl}, \Omega\right)}

\newcommand*{\ldblbrace}{\{\mskip-5mu\{}
\newcommand*{\rdblbrace}{\}\mskip-5mu\}}

\usepackage[backend=biber, style=ieee, url=true]{biblatex}

\AtEveryBibitem{
  \ifentrytype{online}{%
  }{%
    \clearfield{howpublished}%
    \clearfield{url}%
    \clearfield{urldate}%
    \clearfield{issn}%
  }%
}

\pgfplotsset{
  colormap/plasma/.style={%
      /pgfplots/colormap={plasma}{%
        rgb=(0.050383, 0.029803, 0.527975)
        rgb=(0.186213, 0.018803, 0.587228)
        rgb=(0.287076, 0.010855, 0.627295)
        rgb=(0.381047, 0.001814, 0.653068)
        rgb=(0.471457, 0.005678, 0.659897)
        rgb=(0.557243, 0.047331, 0.643443)
        rgb=(0.636008, 0.112092, 0.605205)
        rgb=(0.706178, 0.178437, 0.553657)
        rgb=(0.768090, 0.244817, 0.498465)
        rgb=(0.823132, 0.311261, 0.444806)
        rgb=(0.872303, 0.378774, 0.393355)
        rgb=(0.915471, 0.448807, 0.342890)
        rgb=(0.951344, 0.522850, 0.292275)
        rgb=(0.977856, 0.602051, 0.241387)
        rgb=(0.992541, 0.687030, 0.192170)
        rgb=(0.992505, 0.777967, 0.152855)
        rgb=(0.974443, 0.874622, 0.144061)
        rgb=(0.940015, 0.975158, 0.131326)
  },
},
colormap/magma/.style={%
/pgfplots/colormap={magma}{%
  rgb=(0.001462, 0.000466, 0.013866)
  rgb=(0.035520, 0.028397, 0.125209)
  rgb=(0.102815, 0.063010, 0.257854)
  rgb=(0.191460, 0.064818, 0.396152)
  rgb=(0.291366, 0.064553, 0.475462)
  rgb=(0.384299, 0.097855, 0.501002)
  rgb=(0.475780, 0.134577, 0.507921)
  rgb=(0.569172, 0.167454, 0.504105)
  rgb=(0.664915, 0.198075, 0.488836)
  rgb=(0.761077, 0.231214, 0.460162)
  rgb=(0.852126, 0.276106, 0.418573)
  rgb=(0.925937, 0.346844, 0.374959)
  rgb=(0.969680, 0.446936, 0.360311)
  rgb=(0.989363, 0.557873, 0.391671)
  rgb=(0.996580, 0.668256, 0.456192)
  rgb=(0.996727, 0.776795, 0.541039)
  rgb=(0.992440, 0.884330, 0.640099)
  rgb=(0.987053, 0.991438, 0.749504)
},
},
colormap/plasmaconstant/.style={%
/pgfplots/colormap={plasmaconstant}{%
  rgb=(0.768090, 0.244817, 0.498465)
  rgb=(0.823132, 0.311261, 0.444806)
},
},
}

\pgfplotsset{
  myplotstyle/.style={
    width=0.45\textwidth,
    height=6cm,
    axis line style={thick},
    tick style={thick},
    label style={font=\small},
    tick label style={font=\footnotesize},
    grid=none,
    grid style={dashed,gray!50},
    legend style={font=\scriptsize, draw=none, fill=white},
    cycle list name=exotic,
    every axis plot/.append style={thick}, 
    axis on top=false, 
    tick label style={
      /pgf/number format/use comma=false,
      /pgf/number format/1000 sep={\,}, 
    },
  },
    every axis/.append style={myplotstyle}, 
}

\begin{document}

\title{Surface Contact Approximation for Magneto-Thermal Finite Element Analysis of No-Insulation HTS Coils}

\author{Erik Schnaubelt, Louis Denis, Mariusz Wozniak, Julien Dular, and Arjan Verweij
\thanks{E. Schnaubelt, M. Wozniak, J. Dular, and A. Verweij are with CERN, 1211 Meyrin, Switzerland (e-mail: erik.
schnaubelt@cern.ch). L. Denis and J. Dular are with the University of Liège, 4000 Liège, Belgium.}
\thanks{E. Schnaubelt, M. Wozniak, and J. Dular were partially supported by the CERN High-Field Magnet (HFM) program. L. Denis is a research fellow of the Fonds de la Recherche Scientifique - FNRS.}}

\markboth{}%
{Schnaubelt \MakeLowercase{\textit{et al.}}: SCA Formulation for Analysis of NI Coils}


\maketitle

\begin{abstract}
High-temperature superconducting (HTS) coated conductors (CCs) can be wound into no-insulation (NI) coils, in which electrical current can partially bypass local normal zones via turn-to-turn contact layers (T2TCLs). Accurate magneto-thermal simulation of such coils, therefore, requires an efficient representation of the electrical and thermal behavior of the T2TCLs. This paper introduces a magneto-thermal surface contact approximation (SCA) for finite element analysis of NI HTS coils. The formulation is derived as a special case of the more general thin shell approximation (TSA) by introducing suitable approximations such as negligible tangential surface currents and eddy-current effects inside the T2TCL. The resulting SCA formulation replaces the thin volumetric contact layer with a dedicated surface weak formulation based on the electric contact resistance and thermal contact conductance. In contrast, the TSA formulation requires the definition of electric resistivities and thermal conductivities as well as the thickness of the T2TCL.

The SCA is implemented in the Pancake3D module of the free and open-source Finite Element Quench Simulator (FiQuS). It is verified through transient magneto-thermal simulations of a model NI pancake coil. Numerical results are compared against the established TSA formulation. The results show that the SCA accurately reproduces the relevant electromagnetic and thermal behavior. For the TSA, there is a trade-off between choosing large (potentially unphysical) thicknesses with low resistivities leading to inaccurate results, or small thicknesses with large resistivities making the linear system harder to solve, increasing the computational effort. In contrast, the SCA, thanks to using contact resistances and conductances directly without the necessity to define a thickness, is easy to use and robust. Furthermore, the SCA formulation leads to fewer degrees of freedom than its TSA counterpart. The full software environment, input files, and instructions to reproduce all presented results are publicly available.
\end{abstract}


\section{Introduction}

\IEEEPARstart{N}{o-insulation} (NI) coils~\cite{Hahn2011} wound from high-temperature superconducting (HTS) coated conductors (CCs), i.e., coils without electrical insulation in-between the turns, are an interesting option for high-field solenoid magnets. This is due to the possibility that the current can bypass local normal zones by crossing the turn-to-turn contact layer (T2TCL) \cite{Wang2015}. Therefore, NI coils are commonly characterized by high thermal stability and tolerance to local degradation \cite{Hahn_2016, Wang_2016aa}. Despite increased thermal stability, NI coils are still subject to quenches, see, e.g., \cite{Kim_2017aa, Hahn_2019aa, Suetomi_2021aa}. Furthermore, turn-to-turn bypass currents lead to significant charging characteristic times. Therefore, methods to control the electrical T2TCL are of interest, e.g., by using metal-insulation coils \cite{Lecrevisse_2022} with T2TCL resistances of \SIrange{100}{10000}{\micro \ohm \centi \meter \squared} with several tens of \si{\micro\meter} T2TCL thickness, or coating with thin metallic cladding \cite{Lu_2018} with T2TCL resistances of \SIrange{10}{1000}{\micro \ohm \centi \meter \squared} with \SIrange{0.1}{1}{\micro\meter} T2TCL thickness.

Numerical modeling can support the design of NI coils and their quench protection systems. In this context, significant research effort has been spent to improve modeling techniques for NI coils. A recent overview of different employed numerical modeling techniques is found at \cite[Section 6.1.1]{Schnaubelt_DISS}. Importantly, the numerical modeling tool needs to accurately represent the electrical and thermal T2TCL behavior, accounting for a range of resistivities and thicknesses. 

Recent advancements like the Pancake3D module \cite{Atalay_2024aa} of the free and open-source Finite Element Quench Simulator (FiQuS) \cite{Vitrano_2023aa} allow for comprehensive three-dimensional magneto-thermal simulations of NI coils. Among others, it features a magneto-thermal thin shell approximation (TSA) \cite{Schnaubelt_2023aa, Schnaubelt_2024aa} that reduces the thin volumetric T2TCL into a surface. With the TSA, high-quality meshes can be achieved with fewer mesh elements, leading to fewer degrees of freedom (DoF) \cite[Section 4.3]{Schnaubelt_2023aa}. This significantly simplifies the meshing process and reduces the computational effort compared to a model with classical volumetric T2TCL.

While the TSA outperforms classical volumetric T2TCL models, it suffers from stability issues for large electrical resistivities of the T2TCL. This leads to increased computational time, as shown by the numerical experiments of this work. These stability issues for larger T2TCL resistivities might be similar to those observed when using a full $\vec{H}$ formulation with unphysically high resistivity for non-conducting domains \cite{Lahtinen_2012aa, Dlotko_2019aa}. The numerical experiments also suggest that the stability issue can be mitigated by artificially increasing the thickness of the T2TCL in the TSA solution, thereby reducing the resistivity. However, too large thicknesses may lead to inaccurate results. Hence, the choice of a combination of resistivity and thickness of the T2TCL for the TSA is not straightforward.

To mitigate this issue, this paper proposes a magneto-thermal surface contact approximation (SCA) that is specifically designed for T2TCLs of NI coils. The SCA considers surface electrical contact resistances and thermal contact conductances directly, without the need to specify a thickness. Commonly, the contact quantities are measured directly experimentally without defining the thickness, see, e.g., \cite{Lee_2021aa, Seok_2018aa}. The SCA is therefore easier to use than the TSA.

In this work, the SCA is derived in detail from the more general magneto-thermal TSA by introducing several assumptions specific to NI coils. In particular, the SCA considers the magnetic field strength tangentially continuous across the T2TCL, implying no surface currents along the T2TCL. The SCA is therefore not suitable for highly conductive regions such as the HTS layer. The SCA is verified by comparison against the TSA method for magneto-thermal transient simulations of a model NI pancake coil. The results highlight that the SCA is more robust than the TSA for large electrical resistivity of the T2TCL, leading to fewer solves of the linear system. Furthermore, it is shown that the SCA assumptions are justified for the modeled coil as it reproduces results obtained with the TSA with excellent accuracy. 

The magnetoquasistatic part of the SCA is identical to the formulation used in \cite{Wang_2024aa, Dular_2025aa}, whereas the thermal SCA and the magneto-thermal coupling, the derivation from and verification against the established TSA, and the application to magneto-thermal transients in NI coils are novel. The full software environment used is free and open-source; input files and instructions to recreate all numerical results are publicly available \cite{analysis_repo}.

Section~\ref{sec:model} introduces the mathematical model of the HTS pancake coil and the SCA model, with a detailed derivation given in the appendix. Section~\ref{sec:implementation} discusses implementation details and software used. Section~\ref{sec:numerical} introduces the model problem and presents the numerical results. The paper is concluded in Section~\ref{sec:conclusion}.

\section{Mathematical Model}
\label{sec:model}

In this section, the mathematical formulation of the SCA is presented. It is derived as a special case of the more general magneto-thermal TSA that is discussed in detail in \cite{Schnaubelt_2024aa} and \cite[Chapter 6]{Schnaubelt_2023aa}. First, a summary of the HTS CC model and the magneto-thermal TSA is presented. Afterwards, the SCA model is introduced. To make the introduction of the formulation less abstract, an NI HTS pancake coil is considered as a model problem. Note that the formulation applies to any geometry with highly resistive T2TCL.

\subsection{HTS CC Homogenization}

\begin{figure}[tbh]
    \centering
    \begin{tikzpicture}[scale=0.35, 
    cross/.style={path picture={ 
    \draw[black] (path picture bounding box.south east) -- (path picture bounding box.north west) (path picture bounding box.south west) -- (path picture bounding box.north east);}}
]
    
    \definecolor{copper}{RGB}{116,71,27}
    \definecolor{pancakeHomogenizedCCColor}{RGB}{102, 73, 58}
    \colorlet{TUDa-9d}{olive}
    \colorlet{TUDa-3d}{olive}
    \colorlet{silver}{gray}
    \colorlet{buffer-1}{yellow}
    \colorlet{buffer-2}{white}
    \colorlet{buffer-3}{blue}
    \colorlet{substrate}{gray!30}
    \colorlet{cooling}{red}
    
    \fill[copper, draw=black] (4.5, -4) rectangle (5.5, 2.5); 
    
    \fill[copper, draw=black] (-4, -2.5) rectangle (-2.8, 2.5); 
    \fill[copper, draw=black] (3.8, -2.5) rectangle (4.5, 2.5); 
 
    \fill[copper, draw=black] (-2.5, -2.5) rectangle (-0.7, 2.5);
    \fill[silver, draw=black] (-2.2, -2.2) rectangle (-2.1, 2.2);
    \fill[black, draw=black] (-2.1, -2.2) rectangle (-2, 2.2);
    \fill[buffer-1, draw=black] (-2, -2.2) rectangle (-1.95, 2.2);
    \fill[buffer-2, draw=black] (-1.95, -2.2) rectangle (-1.9, 2.2);
    \fill[buffer-3, draw=black] (-1.9, -2.2) rectangle (-1.85, 2.2);
    \fill[substrate, draw=black] (-1.85, -2.2) rectangle (-1, 2.2);

    \begin{scope}[xshift=2.1cm]
        \fill[copper, draw=black] (-2.5, -2.5) rectangle (-0.7, 2.5);
        \fill[silver, draw=black] (-2.2, -2.2) rectangle (-2.1, 2.2);
        \fill[black, draw=black] (-2.1, -2.2) rectangle (-2, 2.2);
        \fill[buffer-1, draw=black] (-2, -2.2) rectangle (-1.95, 2.2);
        \fill[buffer-2, draw=black] (-1.95, -2.2) rectangle (-1.9, 2.2);
        \fill[buffer-3, draw=black] (-1.9, -2.2) rectangle (-1.85, 2.2);
        \fill[substrate, draw=black] (-1.85, -2.2) rectangle (-1, 2.2);
    \end{scope} 

    \begin{scope}[xshift=4.2cm]
        \fill[copper, draw=black] (-2.5, -2.5) rectangle (-0.7, 2.5);
        \fill[silver, draw=black] (-2.2, -2.2) rectangle (-2.1, 2.2);
        \fill[black, draw=black] (-2.1, -2.2) rectangle (-2, 2.2);
        \fill[buffer-1, draw=black] (-2, -2.2) rectangle (-1.95, 2.2);
        \fill[buffer-2, draw=black] (-1.95, -2.2) rectangle (-1.9, 2.2);
        \fill[buffer-3, draw=black] (-1.9, -2.2) rectangle (-1.85, 2.2);
        \fill[substrate, draw=black] (-1.85, -2.2) rectangle (-1, 2.2);
    \end{scope} 

    \fill[TUDa-9d, draw=black] (-2.8, -2.5) rectangle (-2.5, 2.5); 
    \fill[TUDa-9d, draw=black] (3.5, -2.5) rectangle (3.8, 2.5); 
 
    \fill[TUDa-3d, draw=black] (-0.7, -2.5) rectangle (-0.4, 2.5); 
    \fill[TUDa-3d, draw=black] (1.4, -2.5) rectangle (1.7, 2.5);
    
    \fill[copper, draw=black] (-5, -2.5) rectangle (-4, 4); 

    \draw[->, thick] (6,0) -- (7.5, 0);

    \fill[cooling] (4.5, -4) rectangle (5.5, -3.75);
    \fill[cooling] (-5, 4) rectangle (-4, 3.75);

    \begin{scope}[xshift=13cm]
        \fill[copper, draw=black] (4.5, -4) rectangle (5.5, 2.5); 
        
        \fill[copper, draw=black] (-4, -2.5) rectangle (-2.7, 2.5); 
        \fill[copper, draw=black] (3.7, -2.5) rectangle (4.5, 2.5); 
     
        \fill[pancakeHomogenizedCCColor!50, draw=black] (-2.6, -2.5) rectangle (-0.6, 2.5);
    
        \begin{scope}[xshift=2.1cm]
            \fill[pancakeHomogenizedCCColor!50, draw=black] (-2.6, -2.5) rectangle (-0.6, 2.5);
        \end{scope} 
    
        \begin{scope}[xshift=4.2cm]
            \fill[pancakeHomogenizedCCColor!50, draw=black] (-2.6, -2.5) rectangle (-0.6, 2.5);
        \end{scope} 
        
        \fill[copper, draw=black] (-5, -2.5) rectangle (-4, 4); 

        \fill[cooling] (4.5, -4) rectangle (5.5, -3.75);
        \fill[cooling] (-5, 4) rectangle (-4, 3.75);

        \fill[TUDa-9d, draw=TUDa-9d] (-2.7, -2.5) rectangle (-2.6, 2.5); 
        \fill[TUDa-9d, draw=TUDa-9d] (3.6, -2.5) rectangle (3.7, 2.5); 
     
        \fill[TUDa-3d, draw=TUDa-9d] (-0.6, -2.5) rectangle (-0.5, 2.5); 
        \fill[TUDa-3d, draw=TUDa-9d] (1.5, -2.5) rectangle (1.6, 2.5);

        \draw[black] (-4, -2.5) -- (4.5, -2.5);
        \draw[black] (-4, 2.5) -- (4.5, 2.5);
    \end{scope}

    \draw[magenta, dashed, thick] (-2.5, -2.6) rectangle (-0.7, 2.6);

    \draw[magenta, thick, dashed] (-1.9, -2.6) -- (-3, -3) node[inner sep = 1pt, below, font=\footnotesize]{HTS CC}; 
    
    \draw[thick] (1.55, 1.5) -- (1.55, 3) node[inner sep = 1pt, inner sep = 1pt, above, TUDa-3d, font=\footnotesize]{Volumetric T2TCL}; 
    
    \draw[thick] (5, -2) -- (2.25, -3) node[inner sep = 1pt, below, copper, font=\footnotesize]{Terminal};

    \begin{scope}[xshift=13cm]
        \draw[black, thick] (-1.5, -1.5) -- (-0.5, -3) node[inner sep = 1pt, below, pancakeHomogenizedCCColor!50, font=\footnotesize]{Homogenized HTS CC}; 
        \draw[black, thick] (1.55, 1.5) -- (1.55, 3) node[above, TUDa-3d, inner sep = 1pt, font=\footnotesize]{Surfacial T2TCL}; 
        
        
    \end{scope}

 \end{tikzpicture}
    \caption{Cross-section of a small model NI pancake coil with two turns. The left image shows the coil before homogenization of the HTS CC and introduction of the SCA. The layered structure of the HTS CC is represented by different colors schematically. After homogenization, as shown in the right image, the different layers of the HTS CC are represented by a single homogeneous material. The volumetric contact layer is reduced to a surface contact layer using a dedicated weak formulation introduced in Section \ref{sec:sca_form}. The cooling condition is applied to the electric ports marked in red.}
    \label{fig:homogenization_contact_res}
\end{figure}
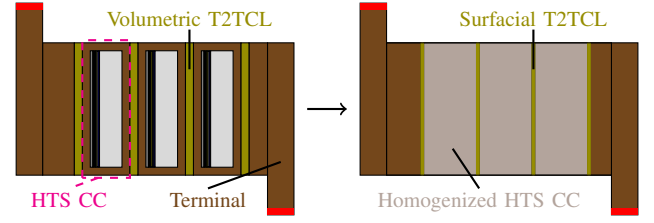

The different layers of the HTS CC are represented by a homogeneous material as shown in Figure~\ref{fig:homogenization_contact_res} and discussed in \cite[Section 6.2]{Schnaubelt_2023aa}. In particular, the homogenization takes the current sharing between superconducting and normal conducting layers into account \cite[Section 6.2.1.1]{Schnaubelt_2024aa}. For the current sharing model, the resistivity of the superconducting layer is given by the power law. In contrast to \cite{Schnaubelt_2024aa}, where an anisotropic electric resistivity and thermal conductivity have been used, both material quantities are assumed to be isotropic here. They are equal to the value of the tensor of \cite{Schnaubelt_2024aa} along the winding direction. That means that both quantities reduce to scalars, which simplifies the problem. This simplification has been made since the focus is on modeling the behavior of the T2TCL and not the HTS CC. More details and comprehensive mathematical expressions for the homogenization of the HTS CC are found in \cite[Section 6.2.3]{Schnaubelt_2023aa}.

\subsection{Magneto-Thermal TSA}

Following \cite{Schnaubelt_2023ab, Schnaubelt_2024aa}, the thin volumetric T2TCL is represented by a surface using a dedicated FE weak formulation that represents the magneto-thermal dynamics across the T2TCL using a TSA. With this approach, meshing the thin layer as a volume can be avoided, such that a high-quality mesh of the coil can be created with fewer mesh elements. This leads to fewer DoF and lower computational effort compared to a model with classical volumetric T2TCL, especially for vanishingly thin T2TCL \cite[Section 4.3]{Schnaubelt_2023aa}. As opposed to \cite{Atalay_2024aa}, no explicit distinction is made between contact layers between two turns of the winding ("winding contact layers") and contact layers between the terminal and winding ("terminal contact layers"). They are all considered by the term T2TCL in this work, for conciseness.

As discussed in the introduction, the idea of the SCA is to simplify the general TSA formulation for the special case of T2TCL with comparably higher resistance than the HTS CC. This leads to a representation of the electrical behavior of the T2TCL in terms of the contact resistance $R_\text{cl} = \rho_\text{cl} \, t_\text{cl}$ in \si{\ohm \meter \squared} instead of $\rho_\text{cl}$, with the electric resistivity $\rho_\text{cl}$ in \si{\ohm \meter} and thickness $t_\text{cl}$ in \si{\meter}. Since the numerical value of the contact resistance $R_\text{cl}$ is smaller than $\rho_\text{cl}$ thanks to the multiplication by $t_\text{cl}$, the resulting linear system of the SCA is expected to be easier to solve than that of the TSA. The numerical experiments of Section~\ref{sec:numerical} showcase this improved robustness of the SCA over the TSA.

\subsection{Weak Magneto-Thermal SCA Formulation}
\label{sec:sca_form}

In this subsection, the SCA formulation is introduced. The comprehensive derivation from the magneto-thermal TSA is given in the appendix for completeness. 

A magneto-thermal transient problem is considered on a computational domain $\Omega \subset \mathbb{R}^3$. The electrically conducting subdomain is referred to as $\Omega_\text{c}$, and its complement is the non-conducting domain $\Omega_\text{c}^\mathsf{c}$. The T2TCL surface is denoted as $\Gamma_\text{cl}$. The volume integral in $\Omega$ of the scalar product of the two arguments is denoted by $(\cdot,\cdot)_\Omega$, with the analogous surface integral on $\Gamma$ denoted as $\left<{\cdot},{\cdot}\right>_{\Gamma}$. For the magnetoquasistatic part of the SCA formulation, an $\vec{H}-\phi$ formulation is used \cite{Pellikka_2013aa}. Its weak formulation reads: find $\vec{H} \in \Hcurl$ such that, $\forall \vec{H}' \in \Hcurlzero$,
\begin{equation}
    \begin{split}
        \left(\partial_t \left(\mu \vec{H} \right), \vec{H}' \right)_{\Omega} +  
        \left( \rho \curl \vec{H}, \curl \vec{H}' \right)_{\Omega_\text{c}}& \\
        + \left<R_\text{cl} \, \curl \vec{H}, \curl \vec{H}'  \right>_{\Gamma_\text{cl}}
        &=  0.
     \end{split} \label{eq:mqs}
\end{equation}

Herein, $\vec{H}$ is the magnetic field strength in \si{\ampere \per \meter}, $\rho$ the electric resistivity in \si{\ohm \meter}, and $\mu$ the magnetic permeability in \si{\henry \per \meter}. The function space $\Hcurl$ is the subspace of $\Hcurlnophi$, i.e., the space of square-integrable functions with square-integrable curl, with vanishing curl in $\Omega_\text{c}^\mathsf{c}$ and current $I$ imposed strongly with edge cohomology basis functions \cite{Pellikka_2013aa}. The space $\Hcurlzero$ is the special case of vanishing current $I = 0$. More details on function spaces and boundary conditions are given in \cite[Section II]{Schnaubelt_2024aa}. 

In formulation~\eqref{eq:mqs}, the electric contact resistance of the T2TCL surface $R_\text{cl}$ is considered by the surface integral on $\Gamma_\text{cl}$ following \cite{Wang_2024aa, Dular_2025aa}. As shown in the appendix, the additional integral is a special case of the magnetoquasistatic TSA of \cite{de-Sousa-Alves_2021aa}. In contrast to the latter, the tangential magnetic field strength is continuous across $\Gamma_\text{cl}$ in the SCA. This implies that surface currents inside $\Gamma_\text{cl}$ cannot be represented. Therefore, the SCA cannot be used to model superconducting layers, while the TSA can represent them accurately \cite{de-Sousa-Alves_2021aa, Alves2022}. For this reason, the name SCA has been chosen to highlight the fact that the approximation is only valid to represent layers with negligible surface currents, such as resistive T2TCL. Removing the discontinuity reduces the number of DoF since the negative and positive sides of the T2TCL surface do not need to be distinguished. Hence, no magnetoquasistatic DoF on the T2TCL is duplicated.

The magnetoquasistatic problem is coupled to the thermal problem: find $T \in \Hgrad$ such that, $\forall T' \in \Hgradzero$ 
\begin{align}
    \label{eq:thermal}
    \begin{split}
        \left(\kappa \nabla T, \nabla T'\right)_{\Omega_\mathrm{c}}
        &+
        \left(C_\mathrm{V}\,\partial_t T, T'\right)_{\Omega_\mathrm{c}} + 
        \left\langle
            Q,
             T' 
        \right\rangle_{\Gamma_\mathrm{t}} + \\
        \left\langle
            K_\mathrm{cl}\,\llbracket T \rrbracket,
            \llbracket T' \rrbracket
        \right\rangle_{\Gamma_\mathrm{cl}}
        &=
        \left(
            \rho\,\lVert\curl \vec{H}\rVert^2,
            T'
        \right)_{\Omega_\mathrm{c}} +
        \\
        & \qquad 
        \left\langle
            R_\mathrm{cl}\,R
            \lVert \curl\vec{H} \rVert^2,
            \ldblbrace T' \rdblbrace
        \right\rangle_{\Gamma_\mathrm{cl}},
    \end{split}
\end{align}
with the thermal conductivity $\kappa$ in \si{\watt \per \meter \per \kelvin}, the volumetric heat capacity $C_\text{V}$ in \si{\joule \per \kelvin \per \meter \cubed} and the temperature $T$ in \si{\kelvin}. For our model problem, the boundary $\Gamma_\mathrm{t}$ consists of the electric ports of the terminals. There, an inhomogeneous Neumann boundary condition imposes a heat flux $Q$ in \si{\watt \per \meter \squared}. The latter models the cooling power of a cryocooler, as explained in more detail in Section~\ref{sec:numerical}. Other cooling configurations, such as adiabatic conditions, could be considered straightforwardly \cite{Schnaubelt_2023aa}. The function space $\Hgrad$ is the space of square-integrable functions with square-integrable gradient. No essential conditions for temperature are considered in the function space, as no Dirichlet boundary conditions are assumed for the temperature. More details on thermal function spaces and boundary conditions are found in \cite{Schnaubelt_2023aa, Schnaubelt_2024aa}.

Compared to a standard weak formulation of the heat diffusion equation (see, e.g., \cite[Section 6.1.3]{Ern_2004aa}), there are two additional surface terms on $\Gamma_\text{cl}$. To recall, the terms are derived from the general magneto-thermal TSA in the appendix. The first one uses the thermal contact conductance $K_\text{cl} = \kappa_\text{cl}/t_\text{cl}$ in \si{\watt \per \kelvin \per \meter \squared} and the jump operator $\llbracket T \rrbracket = T_+ - T_-$. From the latter, it is clear that the temperature is discontinuous, representing thermal gradients across $\Gamma_\text{cl}$. The temperatures $T_+$ and $T_-$ are the temperatures on the two sides of $\Gamma_\text{cl}$, see \cite[Section 3.1]{Schnaubelt_2023aa} for details. The second term describes the Joule loss caused by T2TCL currents. Here, $\ldblbrace T' \rdblbrace = \frac{1}{2} (T'_+ + T'_-)$ is the average operator. The number of DoF for the thermal SCA model is identical to a thermal TSA with the minimum number of one discretization layer $N = 1$ \cite[Section 3.2]{Schnaubelt_2023aa}. However, the SCA has fewer integral terms in the weak formulation.

\subsection{Discretization Details}

The magnetoquasistatic and thermal formulations \eqref{eq:mqs} and \eqref{eq:thermal} are assembled into a single monolithic, i.e., strongly coupled, system. Lowest-order polynomial basis functions are used for discretization in space. Comprehensive details on the used discretization in space are found in \cite[Section 3.2]{Schnaubelt_DISS}. For time discretization, an adaptive implicit Euler scheme is used \cite[Section 3.3]{Schnaubelt_DISS}. The nonlinear equations are linearized using a quasi-Newton scheme that only considers the derivative of the electric field with respect to the current density, i.e., $\partial \vec{E} / \partial \vec{J}$ \cite[Section 6.2.4]{Schnaubelt_DISS}. All other contributions to the Jacobian are neglected for simplicity. Convergence criteria are based on the absolute and relative change of magnetic energy and maximal temperature between iterations of the quasi-Newton scheme and time steps. A direct linear solver is used to solve the comparably ill-conditioned linear systems \cite[Section 3.5]{Schnaubelt_DISS}.

\section{Implementation and Used Software}
\label{sec:implementation}

The SCA formulation has been implemented as part of the Pancake3D module \cite{Atalay_2024aa} of the free and open-source FiQuS tool \cite{Vitrano_2023aa}. FiQuS is part of the Simulation of Transient Effects in Accelerator Magnets (STEAM) framework \cite{Bortot_2018ab, STEAM_website}. FiQuS is implemented in Python and uses Gmsh \cite{Geuzaine_2009ab} for geometry and meshing. In particular, edge cohomology basis functions are computed using the Gmsh plugin \cite{Pellikka_2013aa}. GetDP \cite{Dular_1998aa} is used as the FE kernel.

Linear systems are solved using the Multifrontal Massively Parallel sparse direct Solver (MUMPS) linear solver \cite{Amestoy_2001aa} via the Portable, Extensible Toolkit for Scientific Computation (PETSc) \cite{Balay_1997aa}. All fitted material functions are implemented in the STEAM material library \cite{steam_material_library}. All input files and instructions to reproduce the results are found in \cite{analysis_repo}, using CERNGetDP 2026.4.3 \cite{CERNGetDP_repository}, FiQuS 2026.5.1 \cite{FiQuS_repository}, and STEAM SDK 2026.5.1 \cite{SDK_repository}.

\section{Numerical Results and Discussion}
\label{sec:numerical}

In this section, the SCA formulation is verified against the TSA formulation~\cite{Schnaubelt_2024aa}. The TSA itself has been verified comprehensively against models with volumetrically meshed T2TCLs in~\cite{Schnaubelt_2024aa} and in~\cite[Section 6.4.1]{Schnaubelt_DISS}. Furthermore, the TSA models have been validated against experimental data in~\cite[Section 6.4.2]{Schnaubelt_DISS}. 

The comparison in this work is conducted on a 50-turn single pancake coil, whose geometrical and material parameters are gathered in Table~\ref{tab:coilparams}. The critical current density $J_{\textrm{c}}(T, \lVert \mu \vec{H}\rVert)$ fit is taken from \cite{Succi_2024aa, Wozniak_2024ab}, it does not account for angular dependence for simplicity. The fit is prepared for a Faraday HTS CC using data from \cite{Wimbush_2017aa, HTSDatabase}. A local critical current $I_{\textrm{c}}$ degradation, or $I_{\textrm{c}}$ defect, is introduced in the center of the coil along 0.25 turns. Locally, the nominal $I_{\textrm{c}}$ value of the HTS CC is reduced from \SI{500}{\ampere} to \SI{312.5}{\ampere} in the defect. Note that the FiQuS Pancake3D module can handle any $I_\text{c}$ variation along the HTS CC length \cite{Atalay_2024aa}. The coil is conduction-cooled to a nominal base temperature of \SI{10}{\kelvin} via the copper terminals whose electric ports, represented by the red boundaries in Figure~\ref{fig:homogenization_contact_res}, are cooled through a cryocooler. Its temperature-dependent second-stage cooling power is defined in~\cite[Section A.3]{Schnaubelt_DISS} and models the SHI Cryogenics Group RDE-418D4 4K Cryocooler Series \cite{SHI_2024aa}.

\begin{table}[!tbh]%
    \centering%
    \caption{Geometrical and material parameters of the 50-turn model pancake coil. The reference T2TCL $R_{\textrm{cl}}$ and $K_{\textrm{cl}}$ are chosen in the order of the measured results of \cite[Table III, Heated SMI1]{Lee_2021aa} and \cite[Figure 7]{Seok_2018aa}, respectively.}%
    \label{tab:coilparams}%
    \begin{tabular}{cc}
            \toprule
            Description & Value \\
            \midrule
            Number of turns & 50 \\
            Inner radius of windings & \SI{4}{\centi \meter} \\
            HTS CC thickness x width & \SI{100}{\micro \meter} x \SI{4}{\milli \meter} \\
            HTS CC REBCO thickness & \SI{2.5}{\micro \meter}\\
            HTS CC Hastelloy$^{\text{\textregistered}}$~thickness & \SI{55}{\micro \meter}\\
            HTS CC Copper thickness & \SI{40}{\micro \meter} (RRR: 100) \\
            HTS CC plated Sn thickness & \SI{2.5}{\micro \meter} \\
            $I_{\text{c}}(2.5~\text{T}, 10~\text{K})$ & \SI{500}{\ampere} \\
            REBCO power-law $n$-value & 30 \\
            Reference T2TCL $R_{\textrm{cl}}$ & \SI{5}{\micro \ohm \centi \meter \squared} \\
            Reference T2TCL $K_{\textrm{cl}}$ & \SI{2}{\kilo \watt\per\kelvin\per\meter\squared} \\
            T2TCL $R_{\textrm{cl}}$ range & \SIrange[range-phrase = { -- }, range-units = single]{0.625}{20}{\micro \ohm \centi \meter \squared} \\
            T2TCL $K_{\textrm{cl}}$ range & \SIrange[range-phrase = { -- }, range-units = single]{16}{0.5}{\kilo \watt\per\kelvin\per\meter\squared} \\
            Terminal thickness & 3~mm \\
            Terminal material & Copper (RRR: 100) \\
            Initial temperature & \SI{10}{\kelvin} \\
            Local $I_{\textrm{c}}$ defect & Turn 25 to turn 25.25  \\
            \bottomrule
    \end{tabular}
\end{table}

\begin{figure}[tbh]%
    \centering%
    \input{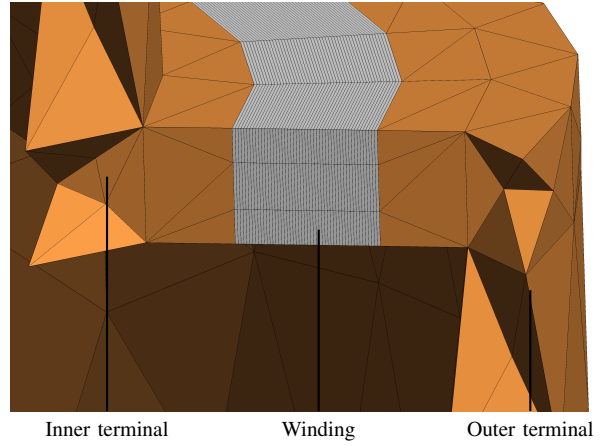}%
    \caption{Finite element mesh of the simulated 50-turn pancake coil, with hexahedral mesh elements within the winding. For better visibility, the air mesh is not represented.}
    \label{fig:mesh}
\end{figure}

The same finite element mesh is used for both the TSA and SCA models for consistent comparison (see Figure~\ref{fig:mesh}). A structured hexahedral mesh is considered in the winding, which is connected to the unstructured tetrahedral air mesh using pyramidal elements. The winding has 36 mesh elements in the azimuthal direction with 3 mesh elements in the axial direction, and a single element per turn in the radial direction. The mesh is chosen rather coarse since the SCA and TSA methods lead to equivalent results independent of the mesh, as long as the assumptions of the SCA are fulfilled. This mesh leads to 45772 and 54949 DoF for the SCA and TSA formulations, respectively. As discussed in Section~\ref{sec:model}, the reduction in problem size is a direct consequence of the simplified treatment of T2TCLs with the SCA.

\subsection{Numerical Verification}
In this subsection, results obtained with the SCA are compared to the results obtained with the TSA. In the SCA, vanishingly thin T2TCL can be modeled with a zero thickness. Indeed, the T2TCL thickness $t_{\textrm{cl}}$ does not appear in the weak formulation described in Section~\ref{sec:sca_form}, which only involves the electric contact resistance $R_{\textrm{cl}}$ and thermal contact conductance $K_{\textrm{cl}}$. On the other hand, the TSA requires a finite $t_{\textrm{cl}}$ value, see~\cite{Schnaubelt_2024aa}. For this reason, the TSA results are computed using a virtual finite thickness, while the geometrical model considers $t_{\textrm{cl}} = 0$. Hereafter, TSA results are shown for multiple virtual T2TCL thicknesses ($R_{\textrm{cl}}$ and $K_{\textrm{cl}}$ values are kept constant).

\begin{figure}
    \centering
    \includegraphics{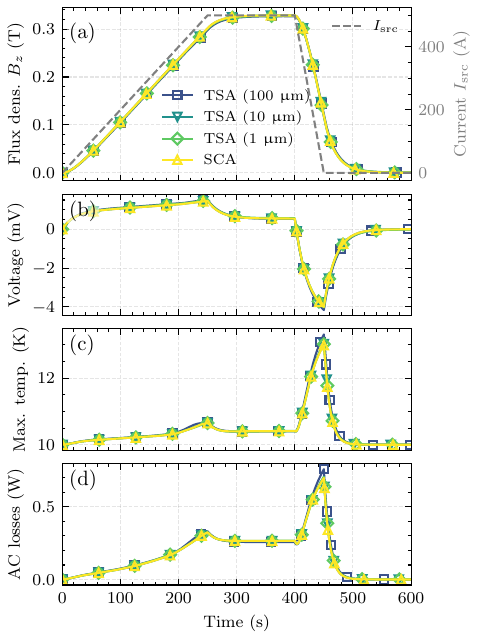}
    \caption{Central axial magnetic flux density $B_z$ and source current $I_{\textrm{src}}$ (a), voltage between copper terminals (b), maximal winding temperature (c), and total AC losses (d) computed with TSA and SCA formulations. TSA results are computed with multiple virtual T2TCL thickness $t_{\textrm{cl}}$ values, indicated in parentheses. Maximal time steps of $1$~s are ensured for consistent comparison between different models. The four figures (a) - (d) share the same legend shown in (a).}
    \label{fig:centralField}
\end{figure}

The applied source current profile is shown in Figure~\ref{fig:centralField}(a), together with the axial central magnetic flux density computed with both TSA and SCA models. The different models are in good agreement, and the typical delay of the magnetic field with respect to the source current is obtained. Similarly, the voltage between the electric ports of the copper terminals, the maximal winding temperature, and the total AC losses are also well reproduced by the SCA model as shown in Figures~\ref{fig:centralField}(b) - (d). Note that the TSA with virtual thickness $t_{\textrm{cl}} = 100$~$\upmu$m overestimates the temperature increase during discharge, as discussed below. The voltage is maximal (in absolute value) during charging and discharging, while it does not decay to zero during the source current plateau due to radial currents bypassing the $I_{\textrm{c}}$ defect and resistive voltage in the copper terminals. The maximal winding temperature and the AC loss curves are very similar, as the former is a direct consequence of the latter. Accordingly, the temperature first rises during charging, before reaching a constant value during the current plateau. The temperature elevation is non-zero during the plateau due to finite AC losses mostly occurring in and next to the $I_{\textrm{c}}$ defect. The temperature then shows its largest elevation during current discharge due to T2TCL currents rapidly dissipating the stored magnetic energy through AC losses.

\begin{figure}
    \centering
    \includegraphics{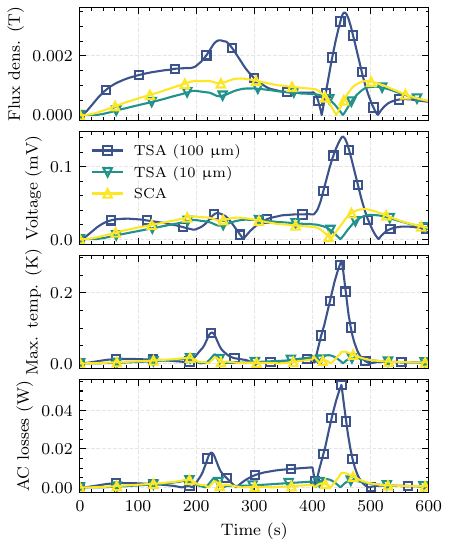}
    \caption{Absolute difference in central axial magnetic flux density (a), voltage between copper terminals (b), maximal winding temperature (c), and total AC losses (d) with respect to TSA results with $t_{\textrm{cl}} = 1$~$\upmu$m. TSA results are computed with multiple virtual T2TCL thickness $t_{\textrm{cl}}$ values, indicated in parentheses. Maximal time steps of $1$~s are ensured for consistent comparison between different models. The four figures (a) - (d) share the same legend shown in (b).}
    \label{fig:errorPlots}
\end{figure}

The absolute difference in numerical results, computed with respect to the TSA model with $t_\textrm{cl} = \SI{1}{\micro \meter}$, is represented in Figure~\ref{fig:errorPlots} for the SCA as well as for the TSA with larger virtual thicknesses. As can be observed, a larger virtual thickness in the TSA can lead to discrepancies in numerical results. Indeed, the TSA with $t_\textrm{cl} = \SI{100}{\micro \meter}$ slightly overestimates the central magnetic flux density as well as the voltage. The discrepancies in maximal winding temperature and in total AC losses are even more pronounced, in particular during transient phenomena, i.e., during charge and discharge of the coil. This can be explained physically by the spurious change in magneto-thermal properties when the TSA contact layer thickness is virtually increased. While its equivalent radial electric resistance is consistent with the reference, its longitudinal electric resistance is strongly decreased by the change in virtual thickness. Similarly, thermal properties are also impacted. As a consequence, increasing the TSA thickness only leads to consistent results in a specific range of virtual thicknesses. On the other hand, the SCA provides consistent results without requiring the choice of such virtual thickness as it relies explicitly on physical contact quantities ($R_{\textrm{cl}}$ and $K_{\textrm{cl}}$) that can be determined experimentally, which highlights its robustness.

\begin{figure}[tbh]
    \centering
    \includegraphics[width=0.45\textwidth]{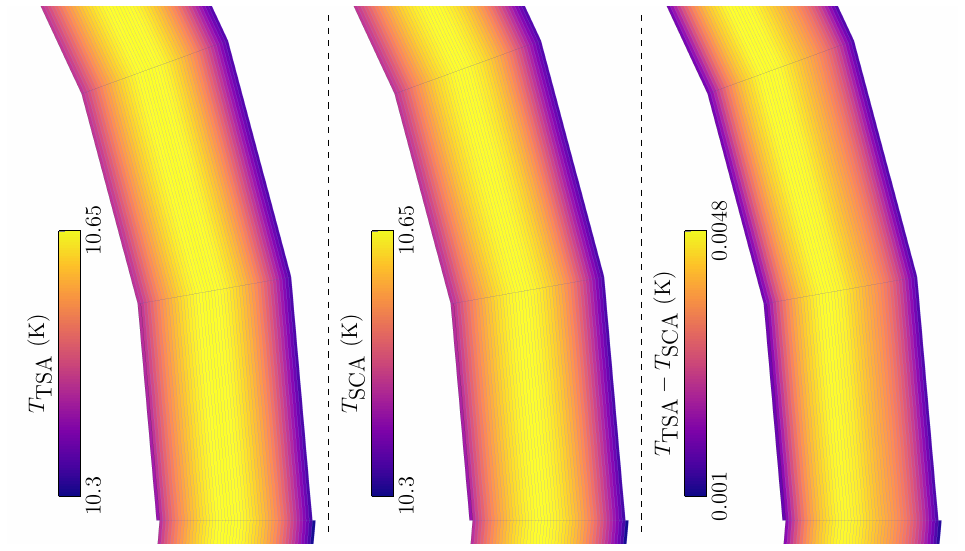}
    \caption{Temperature distribution at $t = 250$~s computed with the TSA and $t_{\textrm{cl}} = 1$~$\upmu$m (left), with the SCA (center), and corresponding absolute difference (right). }
    \label{fig:temp_maps}
\end{figure}

The temperature distribution across the winding is represented in Figure~\ref{fig:temp_maps} at the end of the source current charge: it is maximal at the center of the pancake, near the defect. Again, the SCA reproduces the results from the TSA, with the absolute temperature difference being negligible and smaller than 0.0048~K, which corresponds to less than 1\% of the maximal temperature increase. This verifies the SCA implementation against the TSA and thereby against models with volumetrically meshed T2TCLs.

\subsection{Numerical Performance}
Having verified the consistency of the SCA formulation, this subsection discusses its numerical performance and robustness compared to the TSA formulation. In particular, the electric contact resistance $R_\text{cl}$ is varied, and its influence on the numerical performance of both SCA and TSA models is highlighted. In that context, an inversely-proportional relation between $K_\text{cl}$ and $R_\text{cl}$ is assumed, i.e., the product $K_\text{cl} R_\text{cl}$ is fixed to its reference value (see Table~\ref{tab:coilparams}) for all $R_\text{cl}$ configurations.

Physically, for a source current $I > I_\text{c}$, a larger $R_\text{cl}$ value leads to larger heat dissipation in the T2TCLs near the $I_{\textrm{c}}$ defect, while the simultaneous decrease in $K_\text{cl}$ reduces heat diffusion towards the copper terminals. Therefore, the model pancake can only sustain the source current powering cycle in Figure~\ref{fig:centralField}(a) up to $R_\text{cl} = 20$~\si{\micro\ohm\centi\meter\squared}, while larger $R_\text{cl}$ values induce the thermal runaway of the coil.

\begin{figure}[tbh]
    \centering%
    \includegraphics{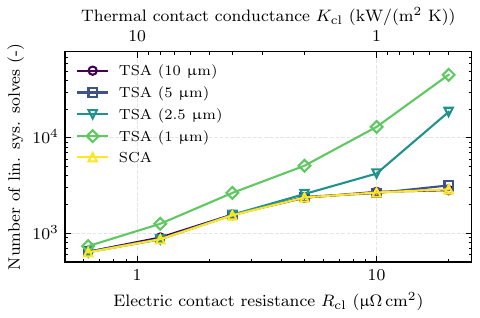}
    \caption{Total number $N_{\textrm{s}}$ of linear system solves required to simulate the powering cycle for various $R_{\textrm{cl}}$ values, with both SCA and TSA formulations. TSA results are computed with multiple virtual T2TCL thickness $t_{\textrm{cl}}$ values, indicated in parentheses.}%
    \label{fig:system_solves}%
\end{figure}

Here, the performance of both formulations is assessed through the total number $N_{\textrm{s}}$ of linear system solves required to simulate the complete powering cycle depicted in Figure~\ref{fig:centralField}(a). In other words, $N_{\textrm{s}}$ represents the total number of quasi-Newton iterations required to perform all time steps of the numerical simulation. It is thus an image of computational cost. The evolution of $N_{\textrm{s}}$ with increasing $R_{\text{cl}}$ values is represented in Figure~\ref{fig:system_solves} for the SCA model and the TSA models with different virtual T2TCL $t_{\textrm{cl}}$ thicknesses. As can be observed, the SCA consistently requires fewer linear system solves than the TSA, which highlights the robustness of the SCA. Moreover, the number of iterations strongly increases for smaller $t_{\textrm{cl}}$ values when increasing the electric contact resistance $R_{\textrm{cl}}$ in the TSA. For $R_{\textrm{cl}} = \SI{20}{\micro\ohm\centi\meter\squared}$, the TSA with $t_{\textrm{cl}} = \SI{1}{\micro \meter}$ requires 15 times more iterations than the SCA. Among other factors, the poorer convergence may be attributed to more ill-conditioned systems obtained after discretization with the TSA than with the SCA. In particular, the TSA solves for the tangential magnetic field discontinuity across each T2TCL, which affects $\vec J = \nabla \times \vec H$ in the HTS CC and the overall quasi-Newton convergence.

To conclude, the SCA requires fewer linear system solves than the TSA, especially for larger T2TCL resistivities. Furthermore, each quasi-Newton iteration of the SCA is less computationally demanding since it is associated with fewer DoF compared to the TSA.

\section{Conclusion}
\label{sec:conclusion}

This paper introduced a magneto-thermal SCA for the efficient FE analysis of NI HTS pancake coils. The formulation was derived as a special case of the more general magneto-thermal TSA. To this end, assumptions were introduced that are physically justified for T2TCLs with comparatively high electrical resistance and negligible surface currents. In particular, the SCA replaces the volumetric representation of the T2TCL by a dedicated surface formulation based on the contact resistance and thermal contact conductance.

The proposed formulation was implemented in the Pancake3D module of the free and open-source FiQuS and verified through transient magneto-thermal simulations of NI pancake coils. The numerical results demonstrated that the SCA reproduces the relevant electromagnetic and thermal behavior of the TSA while significantly improving numerical robustness for larger contact resistivities. While the robustness issue of the TSA can be overcome by choosing appropriate increased virtual thicknesses, the choice of thickness is not straightforward, and a trade-off between accuracy and computational effort must be found. In contrast, the SCA consistently needs fewer or a similar number of linear system solves than the TSA and is easier to use. The results further confirmed that neglecting tangential surface currents and eddy-current effects inside the T2TCL has a negligible impact on the simulation results for the NI coil considered. The presented SCA therefore provides an efficient and robust simplification of the TSA for the modeling of resistive T2TCLs in NI HTS coils.

\appendix

In this section, the SCA is derived as a special case of the magneto-thermal TSA of \cite{Schnaubelt_2024aa}. We start by recalling some symbols from \cite[Section 4.1]{Schnaubelt_DISS}. The number of TSA discretization layers is denoted as $N$, the magnetic field strength on the two sides of the T2TCL are denoted as $\vec{H}_+$ and $\vec{H}_-$, and $\vec{H}_{\text{t},j}$  is the tangential magnetic field strength on the $j$-th TSA layer. The normal unit vector evaluated on $\Gamma_{\textrm{cl}}$ is denoted as $\vec{n}$. The TSA leads to the definition of one-dimensional FE material matrices $M_{lj,\rho}^{(k)}$, $S_{lj,\rho}^{(k)}$, and $ M_{lj,\mu}^{(k)}$ \cite{Schnaubelt_2023ab}. The definition of the subset of them that is relevant for the SCA is given later in this section. From \cite[Equations (4.4), (4.6), (4.12) and (4.14)]{Schnaubelt_DISS}, the TSA surface contributions for the magnetoquasistatic case are given by
\begin{equation}
    \begin{alignedat}{2}
           \mathcal{A} = & \big<\vec{n} \times \left( \rho \curl \vec{H}_+\right), \vec{H}'_+ \big>_{\Gamma_\text{cl}}
           -  \\
           &  \big< \vec{n} \times \left( \rho \curl \vec{H}_- \right), \vec{H}'_- \big>_{\Gamma_\text{cl}} \\
            = & \sum_{k=1}^{N} \sum_{j=k-1}^k \sum_{l=k-1}^k \Bigg[ \left< M_{lj,\rho}^{(k)} \curl \vec{H}_{\text{t},j}, \curl \vec{H}'_{\text{t},l} \right> _{\Gamma_\text{cl}} 
            +\\
           & \left< S_{lj,\rho}^{(k)} \vec{H}_{\text{t},j}  , \vec{H}'_{\text{t},l} \right>_{\Gamma_\text{cl}}+
            \left< M_{lj,\mu}^{(k)} \partial_t \vec{H}_{\text{t},j} ,  \vec{H}'_{\text{t},l} \right> _{\Gamma_\text{cl}} 
            \Bigg ].
    \end{alignedat}
\end{equation}
The symbol $\mathcal{A}$ is introduced to shorten the notation. The SCA is derived by introducing several assumptions. First, the analysis is restricted to one TSA discretization layer, i.e., $N = 1$. Second, the tangential magnetic field strength is assumed to be continuous, i.e., $ \vec{H}_{\text{t}} =  \vec{H}_{\text{t},0} =  \vec{H}_{\text{t},1}$ and $ \vec{H}'_{\text{t}} = \vec{H}'_{\text{t},0} =  \vec{H}'_{\text{t},1}$. This implies that there are no surface currents in the T2TCL \cite[Section 3.6]{Klingbeil_2018aa}. The latter is a reasonable assumption for T2TCL with comparably higher resistance than the HTS CC. Third, eddy currents induced by the time variation of the magnetic field are neglected in the T2TCL, i.e., $\partial_t \vec{H}_{\text{t},j} = 0$, $\forall j$. Again, this is a reasonable assumption given the T2TCL's increased resistance relative to the HTS CC. With these three assumptions, the TSA contributions are simplified to
\begin{equation}
    \begin{alignedat}{2}
           \mathcal{A} = \sum_{j=0}^{1} \sum_{l=0}^{1}\Bigg[
            &\left< M_{lj,\rho}^{(1)} \curl \vec{H}_{\text{t}}, \curl \vec{H}'_{\text{t}} \right> _{\Gamma_\text{cl}} 
            \\+
           &\left< S_{lj,\rho}^{(1)} \vec{H}_{\text{t}}  , \vec{H}'_{\text{t}} \right>_{\Gamma_\text{cl}}
            \Bigg ].
    \end{alignedat} \label{eq:sum_formulation}
\end{equation}
The fourth assumption is that the resistivity $\rho_\text{cl}$ is constant along the thickness $w$ in the T2TCL. In this case, the material matrices can be explicitly evaluated \cite{de-Sousa-Alves_2021aa}. For the mass matrix of the resistivity \cite[Equation C.26]{Schnaubelt_DISS}, we get
\begin{equation}
          M_{lj,\rho}^{(1)} 
          = \int_{w_0}^{w_1} \rho_\text{cl}  N_l N_j \, \text{d}w 
          = \frac{\rho_\text{cl}  t_\text{cl}}{6} 
            \begin{cases}
                2  & \text{for }l = j,\\
                1 & \text{for } l \neq j.
            \end{cases}
\end{equation}
Herein, $N_i$ denotes the $i$-th TSA basis function along the thickness. In this work, first-order Lagrange basis functions are used as explained in \cite[Section 3.2]{Schnaubelt_2023aa}. The sum of the mass matrix entries, as apparent in Equation~\eqref{eq:sum_formulation}, is
\begin{equation}
     \sum_{j=0}^{1} \sum_{l=0}^{1} M_{lj,\rho}^{(1)} = \rho_\text{cl}  t_\text{cl} = R_\text{cl}.
\end{equation}
The stiffness matrix of the stiffness resistivity evaluates to
\begin{equation}
    \begin{split}
          S_{lj,\rho}^{(1)} = \int_{w_0}^{w_1} \rho_\text{cl} \, \partial_w N_l \, \partial_w N_j \, \text{d}w 
              = \frac{\rho_\text{cl}}{t_\text{cl}^2} \begin{cases}
                1 &\text{for }l = j,\\
                -1  & \text{for } l \neq j.
            \end{cases}
    \end{split}
\end{equation}
The sum of matrix entries considered in Equation~\eqref{eq:sum_formulation} vanishes, i.e.,
\begin{equation}
     \sum_{j=0}^{1} \sum_{l=0}^{1} S_{lj,\rho}^{(1)} = 0.
\end{equation}
Therefore, the second integral of Equation~\eqref{eq:sum_formulation} is zero. In summary, the additional TSA term reads as in Equation~\eqref{eq:mqs}, i.e.,
\begin{equation}
    \begin{split}
       \mathcal{A} &= \left< R_\text{cl} \curl \vec{H}_{\text{t}}, \curl \vec{H}'_{\text{t}} \right> _{\Gamma_\text{cl}} \\
        &= \left< R_\text{cl} \curl \vec{H}, \curl \vec{H}' \right> _{\Gamma_\text{cl}}.
    \end{split}
\end{equation}

For the heat diffusion equation, we introduce the temperature of the $j$-th TSA layer $T_j$ and the material matrices $S_{lj,\kappa}^{(k)}$, $M_{lj,\kappa}^{(k)}$ and $M_{lj,C_\text{V}}^{(k)}$, $f_{l,Q}$ \cite{Schnaubelt_2023aa}, again defined later. The additional TSA surface integrals read \cite[Equation 9]{Schnaubelt_2023aa}
\begin{equation}
    \begin{alignedat}{2}
           \mathcal{B} &= \big<\vec{n} \cdot \left(\kappa \gradient T_+\right), T'_+ \big>_{\Gamma_\text{cl}}
           - 
           \left< \vec{n} \cdot \left(\kappa \gradient T_-\right>, T'_- \right>_{\Gamma_\text{cl}} \\
           & = \sum_{k=1}^{N} \Bigg \{ \sum_{j=k-1}^k \sum_{l=k-1}^k \Bigg[ 
           \left< S_{lj,\kappa}^{(k)} T_j  , T'_l \right>_{\Gamma_\text{cl}} + \\
           & \phantom{= \sum_{k=1}^{N}} \left< M_{lj,\kappa}^{(k)} \gradient T_j, \gradient T'_l \right> _{\Gamma_\text{cl}} +
            \left< M_{lj,C_\text{V}}^{(k)} \partial_t T_j , T'_l \right> _{\Gamma_\text{cl}} \Bigg ]  - \\
           & \phantom{= \sum_{k=1}^{N}} \sum_{l=k-1}^k \left< f_{l,Q}^{(k)} , T'_l \right>_{\Gamma_\text{cl}} \Bigg \}. 
    \end{alignedat}
\end{equation}
Analogous to the magnetoquasistatic problem, the symbol $\mathcal{B}$ is introduced as an abbreviation. Again, three assumptions simplify the thermal TSA to the thermal SCA. First, we again set $N = 1$. Second, the heat capacity of the thin T2TCL is neglected, i.e., $M_{lj,C_\text{V}}^{(k)} = 0$, $\forall l,\forall j$. This assumption can be justified by the negligible thermal mass of the vanishingly thin contact layer. Third, the tangential heat flux within the T2TCL is neglected, i.e., $\kappa \gradient T_j = 0$, $\forall j$. This assumption is commonly justified, given the comparably low thermal conductivity of the T2TCL compared to the HTS CC. With these assumptions, the TSA surface integrals simplify to
\begin{equation}
     \mathcal{B} = 
      \sum_{j=0}^1 \sum_{l=0}^1
      \left <S_{lj,\kappa}^{(1)} T_j, T'_l  \right>_{\Gamma_\text{cl}} - \sum_{l=0}^1 \left <f_{l,Q}^{(1)}, T'_l  \right>_{\Gamma_\text{cl}}. 
      \label{eq:simplified_thermal}
\end{equation}
Assuming that the thermal conductivity $\kappa$ is constant along the T2TCL thickness $w$, the stiffness matrix of $\kappa$ reads
\begin{equation}
      S_{lj,\kappa}^{(1)} = \int_{w_0}^{w_1} \kappa \, \partial_w N_l \, \partial_w N_j \, \text{d}w
         = K_\text{cl} \begin{cases}
            1& \text{for }l = j,\\
            -1  & \text{for } l \neq j.
        \end{cases}
\end{equation}
In contrast to the tangential magnetic field strength of the magnetoquasistatic problem, the temperature is discontinuous across the T2TCL. Hence, the corresponding term in Equation~\eqref{eq:simplified_thermal} does not vanish as opposed to the magnetoquasistatic SCA. Since $\rho_\text{cl}$ is constant along $w$, the Joule loss $Q = \rho_\text{cl} \lVert \curl \vec{H} \rVert^2$ is also constant along $w$. Therefore, the right-hand side power density reads
\begin{equation}
    \begin{split}
          f_{l,Q}^{(1)} &= \int_{w_0}^{w_1} Q N_l \, \text{d}w \\
            & =\frac{\rho_\text{cl}}{t_\text{cl}} \lVert \curl \vec{H} \rVert^2 \begin{cases}
                \int_{w_0}^{w_1} w_1 - w \, \text{d}w & \text{for }l = 0\\
                \int_{w_0}^{w_1} w - w_0 \, \text{d}w & \text{for } l = 1
            \end{cases} \\ 
             & = \frac{1}{2} R_\text{cl} \lVert \curl \vec{H} \rVert^2 \quad \text{for } l \in \{0,1\}.
    \end{split}
\end{equation}
Next, we remind ourselves that  $T_0 = T_-$, $T'_0 = T'_-$, $T_1 = T_+$, $T'_1 = T'_+$. Using the jump and average operator, the above derivations can be summarized to find the thermal SCA terms as in formulation \eqref{eq:thermal}, i.e.,
\begin{equation}
     \mathcal{B} = \left<K_\text{cl} \llbracket T\rrbracket, \llbracket T' \rrbracket \right>_{\Gamma_\text{cl}}
            - \left< R_\text{cl} \, \lVert \curl \vec{H} \rVert^2, \ldblbrace T' \rdblbrace \right>_{\Gamma_\text{cl}}.
\end{equation}

\printbibliography

\end{document}